\documentclass[a4paper,11pt]{article}
\usepackage{pos}

\usepackage[symbol]{footmisc}
\usepackage[english]{babel}
\usepackage{graphicx}
\usepackage{graphics}
\usepackage{braket}
\usepackage{bbold}
\usepackage{amsmath}
\usepackage{nicefrac}
\usepackage{dcolumn}
\usepackage{bm}
\usepackage{slashed}
\usepackage{datetime}
\usepackage{mciteplus}
\usepackage{multirow}
\usepackage{siunitx}
\usepackage{booktabs}
\usepackage{color, soul}
\usepackage[usenames,dvipsnames]{xcolor}
\usepackage{float}
\usepackage[utf8]{inputenc}
\usepackage[normalem]{ulem}
\usepackage{mathtools}
\usepackage{setspace}
\usepackage{comment}
\renewcommand{\thefootnote}{\fnsymbol{footnote}}

\newcommand{\LQCD}{\Lambda_{\rm QCD}}

\newcommand{\NLLm}{{\rm NLL/NLO^-}}

\newcommand{\DY}{\Delta Y}

\newcommand{{\HFNRevo}}{\tt HF-NRevo}

\title{$Z$-plus-jet Production with JETHAD-DYnamis: \\ Angular Structure, Polarization, Fixed-Order Matching}
\ShortTitle{$Z$-plus-jet Production with JETHAD-DYnamis}

\author*[a]{Francesco Giovanni Celiberto}
\author[a]{Francesca Lonigro}

\affiliation[a]{Departamento de Física y Matemáticas, Universidad de Alcalá (UAH), Campus Universitario, \\ Alcalá de Henares, E-28805, Madrid, Spain}

\emailAdd{francesco.celiberto@uah.es}
\emailAdd{francesca.lonigro@uah.es}

\abstract{Recent NLO analyses of inclusive Higgs production indicate that high-energy resummation corrections can reach the 10\% level, highlighting the growing impact of high-energy QCD dynamics on electroweak observables from LHC to FCC energies. 
We extend this precision program to Drell-Yan plus jet production by deriving high-energy-resummed predictions for rapidity and transverse-momentum distributions of electroweak bosons. 
Fixed-order NLO accuracy is consistently combined with next-to-leading energy-logarithmic resummation (NLL/NLO). 
We outline a JETHAD-DYnamis--POWHEG matching strategy that retains polarization and angular correlations while enabling realistic lepton-level kinematics. 
This setup provides the first high-energy-resummed description of rapidity-separated electroweak-boson plus jet final states. 
The resulting framework offers a precision-oriented description of Drell-Yan observables for present LHC measurements and the High-Luminosity LHC, where percent-level theoretical control will become increasingly important.}

\FullConference{The 33rd International Workshop on Deep Inelastic Scattering and Related Subjects (DIS2026)\\
4-8 May 2026\\
Bologna, Italy\\}

\begin{document}
\maketitle

\section{High-energy resummation at the precision frontier}
\label{sec:introduction}

Precision measurements of electroweak processes at present and future hadron colliders, from the LHC to prospective FCC energies~\cite{FCC:2025lpp,FCC:2025uan,FCC:2025jtd}, increasingly probe configurations in which the collision energy is much larger than the characteristic hard scales,
$\sqrt{s}\gg\{\mu_i\}\gg\LQCD$.
In this semi-hard regime, energy-enhanced logarithms can become numerically relevant and require all-order resummation.
The Balitsky-Fadin-Kuraev-Lipatov (BFKL) framework~\cite{Fadin:1975cb,Balitsky:1978ic} provides their systematic treatment at leading- and next-to-leading-logarithmic accuracy, while simultaneously opening sensitivity to the low-$x$ gluon sector of the proton~\cite{Bacchetta:2020vty,Bacchetta:2024fci,Celiberto:2021zww,Amoroso:2022eow,Bolognino:2018rhb,Bolognino:2021niq,Hentschinski:2022xnd,Celiberto:2019slj}.
A particularly clean realization is obtained when two hard objects are separated by a large rapidity interval $\DY$.
The resulting multi-Regge kinematics enhances high-energy radiation while retaining perturbatively large transverse scales.
Within hybrid factorization (HyF)~\cite{Celiberto:2020tmb,Bolognino:2021mrc}, the cross section is organized in terms of two process-dependent emission functions connected by the universal NLL BFKL Green's function.
When both emission coefficients are available at NLO, this construction reaches NLL/NLO accuracy; when only one is known beyond LO, the mixed realization $\NLLm$ retains the full NLL Green's function.
Related high-energy formulations for single-inclusive reactions can be found in~\cite{vanHameren:2022mtk,Bonvini:2018ixe,Silvetti:2022hyc}.
This program has been tested in a broad range of semi-inclusive channels, including Mueller-Navelet jets~\cite{Ducloue:2013hia,Celiberto:2015yba,Celiberto:2016ygs,Celiberto:2017ius,Celiberto:2022gji}, forward Drell-Yan production~\cite{Celiberto:2018muu,Golec-Biernat:2018kem}, identified hadrons~\cite{Celiberto:2016hae,Celiberto:2017ptm,Bolognino:2018oth,Celiberto:2020rxb,Celiberto:2022kxx,Celiberto:2017nyx,Bolognino:2019yls,Celiberto:2021dzy,Celiberto:2021fdp,Celiberto:2022zdg,Celiberto:2022keu,Celiberto:2024omj,Feng:2022inv,Celiberto:2022grc}, quarkonia~\cite{Boussarie:2017oae,Celiberto:2022dyf,Celiberto:2023fzz,Celiberto:2022grc}, and rare or exotic states~\cite{Celiberto:2025ogy,Celiberto:2026qiz,Celiberto:2023rzw,Celiberto:2024mab,Celiberto:2024mrq,Celiberto:2024beg,Celiberto:2025dfe,Celiberto:2025ziy,Celiberto:2026kks,Celiberto:2025ipt,Celiberto:2026rdk,Celiberto:2026ooh}.
For precision electroweak phenomenology, Higgs-plus-jet production provides a natural benchmark.
High-energy-resummed studies at LHC and FCC energies~\cite{Celiberto:2020tmb,Celiberto:2023rtu}, together with complementary fixed-order NNLO~\cite{Chen:2014gva,Boughezal:2015dra,Dawson:2022zbb} and transverse-momentum-resummed calculations~\cite{Monni:2019yyr}, show that different logarithmic sectors can have sizeable phenomenological impact.
Recent progress toward extending the high-energy description beyond the present NLL/NLO frontier is provided by the two-loop Higgs impact factor in the Regge limit~\cite{DelDuca:2025vux}.
These developments motivate a controlled matching between fixed-order and high-energy-resummed predictions.
The next precision target is Drell-Yan plus jet production, and in particular $Z$-boson production with a rapidity-separated jet.
Besides its central role in PDF fits and electroweak measurements, the leptonic decay of the $Z$ boson provides direct access to polarization and angular observables at the experimentally reconstructed level.
This makes $Z$-plus-jet production a particularly clean channel in which to assess high-energy-resummation effects under realistic collider conditions.
Within the JETHAD ecosystem, DYnamis provides the dedicated supermodule for Drell-Yan final states, retaining the angular structure, polarization information, and lepton-level kinematics required for this precision program.

\section{High-energy matching for Higgs-plus-jet production}
\label{sec:matching}

\begin{figure*}[!t]
\centering

\includegraphics[scale=0.37,clip]{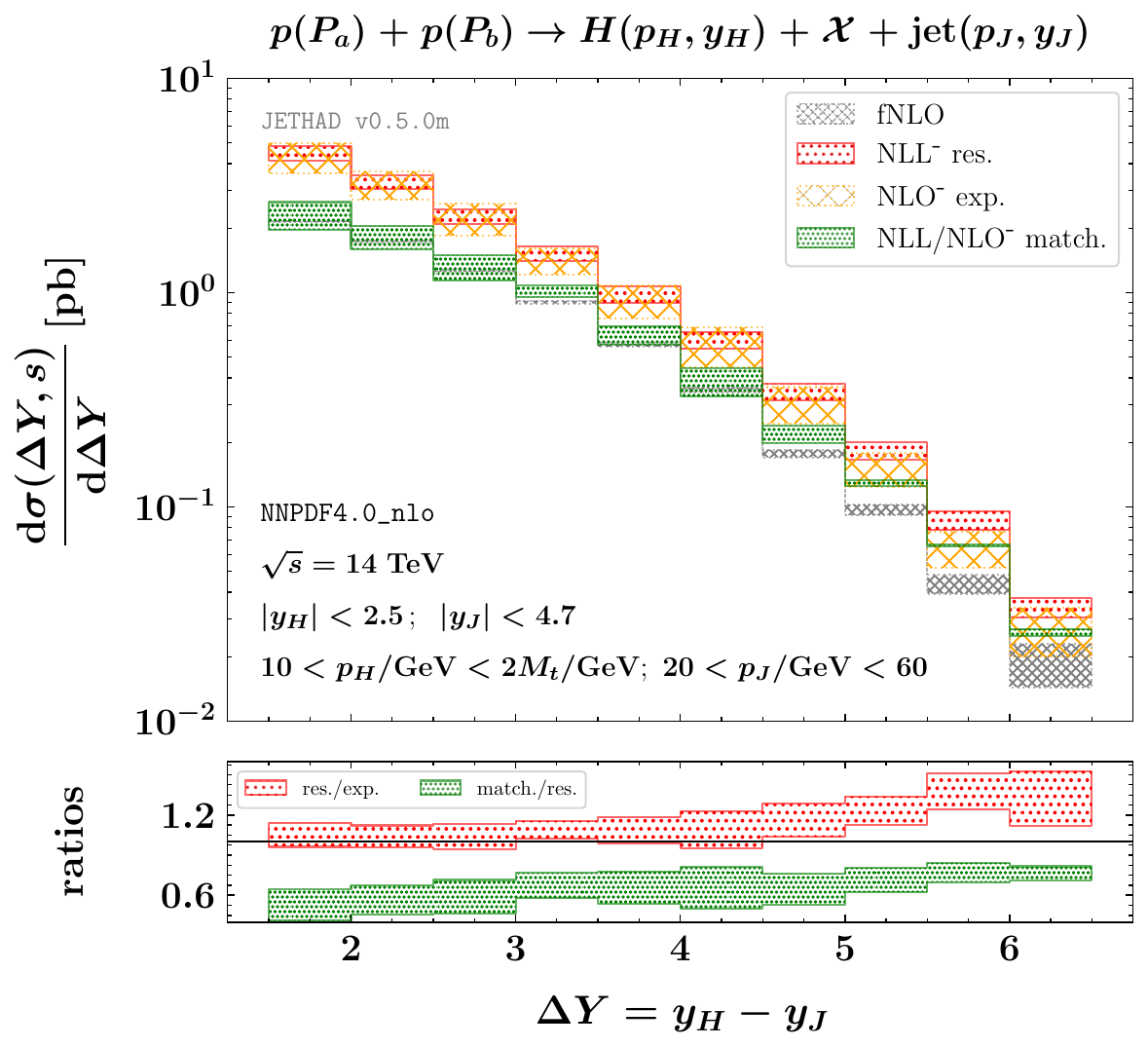}
\hspace{0.00cm}
\includegraphics[scale=0.37,clip]{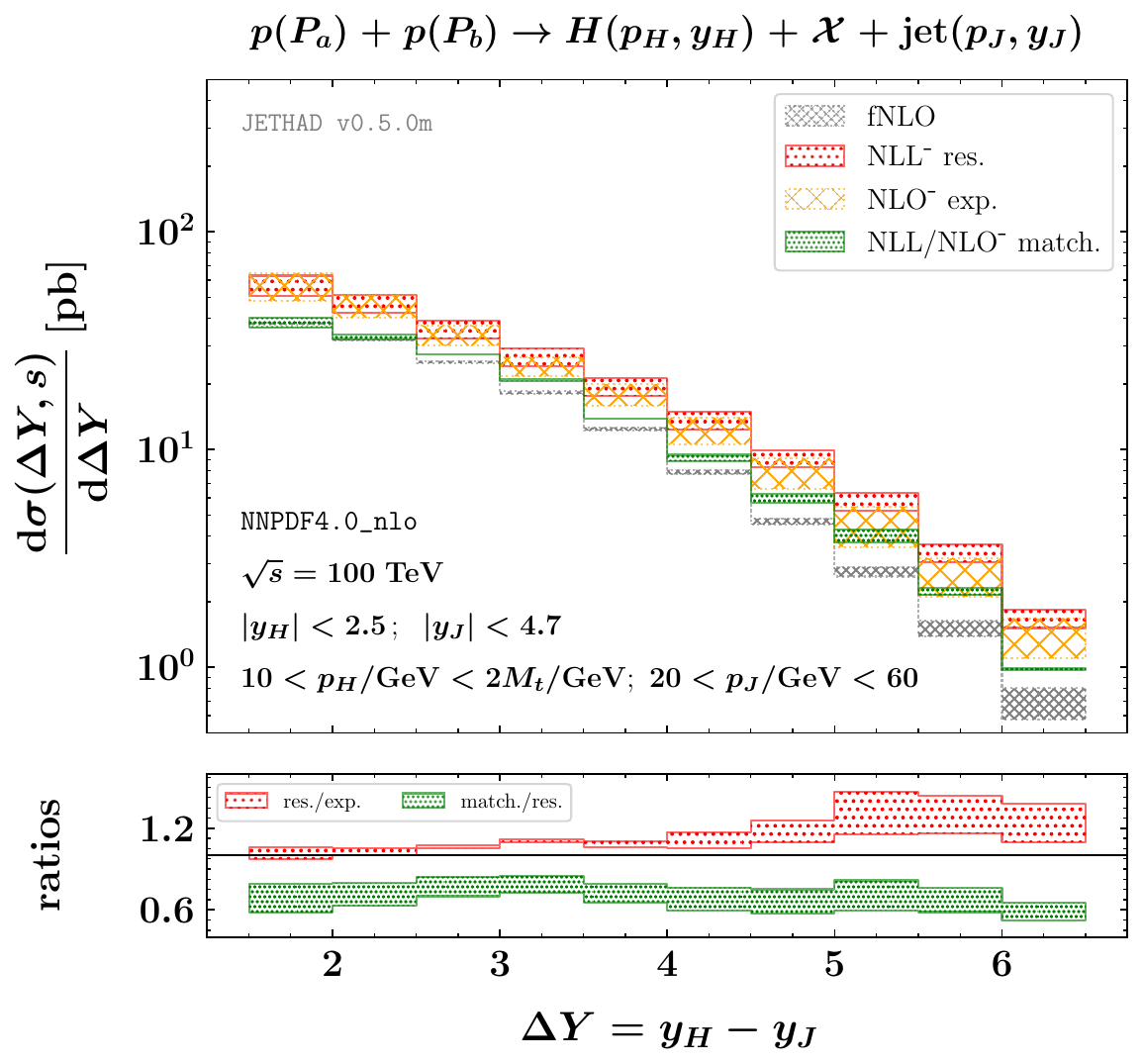}

\caption{Higgs-plus-jet $\DY$ distributions at the 14~TeV LHC (left) and the nominal 100~TeV FCC (right).
Uncertainty bands correspond to $\mu_{R,F}$ scale variations in the range $1<C_{\mu}<2$, while text boxes specify the adopted kinematic cuts.
}

\label{fig:I}
\end{figure*}

\begin{figure*}[!t]
\centering

\includegraphics [scale=0.36,clip]{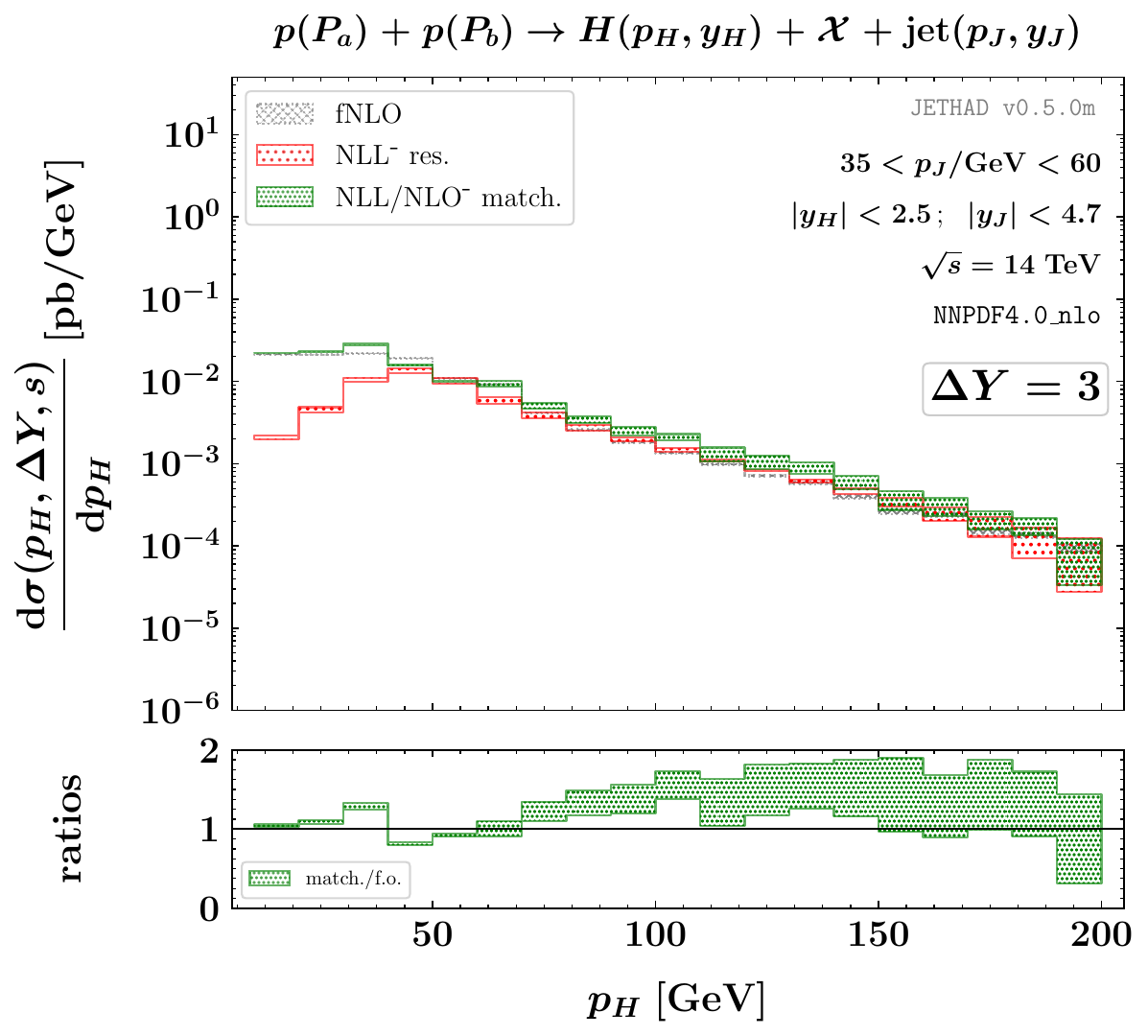}
\hspace{0.00cm}
\includegraphics[scale=0.36,clip]{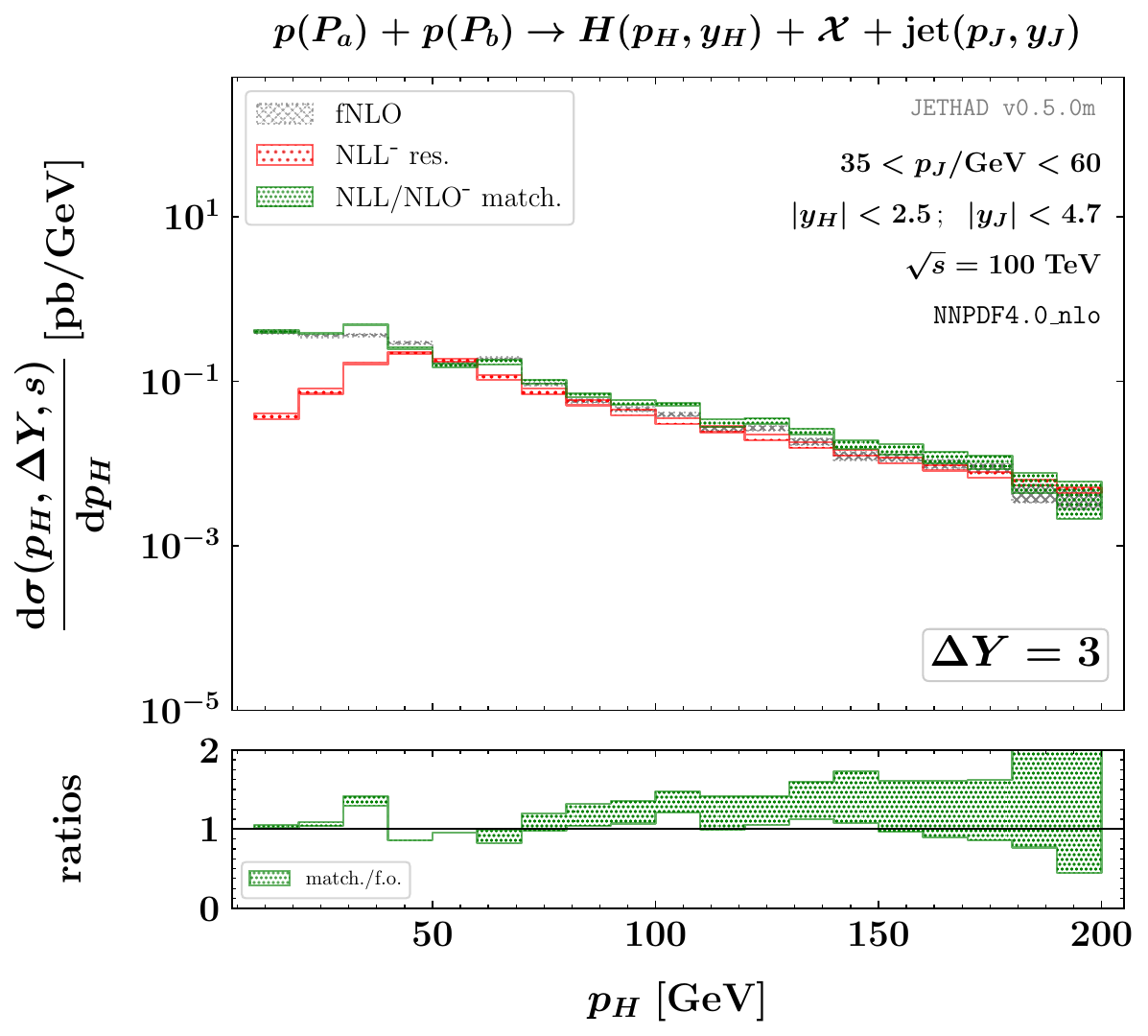}

\caption{Higgs transverse-momentum distributions in Higgs-plus-jet production at the 14~TeV LHC (left) and the nominal 100~TeV FCC (right).
Uncertainty bands correspond to $\mu_{R,F}$ scale variations in the range $1<C_{\mu}<2$, while text boxes specify the adopted kinematic cuts.
}

\label{fig:pT}
\end{figure*}

Previous HyF studies of Higgs-plus-jet production at LHC~\cite{Celiberto:2020tmb} and FCC~\cite{Celiberto:2023rtu} energies revealed a remarkable stability of high-energy-resummed predictions against radiative corrections and scale variations.
At the same time, comparisons with pure fixed-order calculations exposed a non-negligible separation between the two descriptions, particularly in the intermediate transverse-momentum region where their domains of applicability overlap.
This observation motivates a direct matching of fixed-order and high-energy-resummed calculations.

We therefore combine NLO matrix elements with NLL high-energy resummation through an additive matching prescription.
The central requirement is the removal of contributions simultaneously contained in the fixed-order calculation and in the perturbative expansion of the resummed result.
Subtracting this overlap before adding the resummed component preserves the fixed-order content while supplementing it with genuinely all-order energy-logarithmic contributions.

The complete NLO Higgs emission function~\cite{Hentschinski:2020tbi,Celiberto:2022fgx,Nefedov:2019mrg,Celiberto:2024bfu,Celiberto:2026_Higgs-hadron_NLL-NLO} is not yet implemented in JETHAD~\cite{Celiberto:2020wpk,Celiberto:2022rfj,Celiberto:2023fzz,Celiberto:2024mrq,Celiberto:2024swu}.
We consequently employ the mixed $\NLLm$ realization, in which one emission function is evaluated at NLO and the other at LO, both being connected through the full NLL Green's function.
The matching architecture and its POWHEG+JETHAD implementation are discussed in Refs.~\cite{Celiberto:2023uuk,Celiberto:2023eba,Celiberto:2023nym,Celiberto:2024mdt}.
Its essential structure can be summarized as

\begin{equation}
\label{eq:matching}
\begin{split}
 \hspace{-0.255cm}
 \underbrace{{\rm d}\sigma^{{{\rm NLL/NLO}}^{\boldsymbol{-}}}(\Delta Y, \varphi, s)}_{\text{\colorbox{OliveGreen}{\textbf{\textcolor{white}{NLL/NLO$^{\boldsymbol{-}}$}}} {\tt POWHEG+JETHAD}}} 
 = 
 \underbrace{{\rm d}\sigma^{\rm NLO}(\Delta Y, \varphi, s)}_{\text{\colorbox{gray}{\textcolor{white}{\textbf{NLO}}} {\tt POWHEG} w/o PS}}
 +\; 
 \underbrace{\underbrace{{\rm d}\sigma^{{{\rm NLL}}^{\boldsymbol{-}}}(\Delta Y, \varphi, s)}_{\text{\colorbox{red}{\textbf{\textcolor{white}{NLL$^{\boldsymbol{-}}$ resum}}} (HyF)}}
 \;-\; 
 \underbrace{\Delta{\rm d}\sigma^{{{\rm NLL/NLO}}^{\boldsymbol{-}}}(\Delta Y, \varphi, s)}_{\text{\colorbox{orange}{\textbf{NLL$^{\boldsymbol{-}}$ expanded}} at NLO}}}_{\text{\colorbox{NavyBlue}{\textbf{\textcolor{white}{NLL$^{\boldsymbol{-}}$}}} {\tt JETHAD} w/o NLO$^{\boldsymbol{-}}$ double counting}}
 \,.
\end{split}
\end{equation}

Equation~\eqref{eq:matching} separates the matched prediction into an NLO fixed-order baseline, generated with POWHEG~\cite{Hamilton:2012rf,Bagnaschi:2023rbx,Banfi:2023mhz} without parton-shower effects~\cite{Alioli:2022dkj,vanBeekveld:2022zhl,FerrarioRavasio:2023kyg}, and a high-energy correction supplied by JETHAD.
The latter is obtained by subtracting the NLO expansion of the NLL$^-$ result from its fully resummed counterpart, thereby removing double counting.
The color coding adopted in Figs.~\ref{fig:I} and~\ref{fig:pT} follows this decomposition: gray identifies the fixed-order result, red the NLL$^-$ resummed contribution, orange its NLO expansion, and green the final NLL/NLO$^-$ matched prediction.

Extending the preliminary analyses of Refs.~\cite{Celiberto:2023uuk,Celiberto:2023eba,Celiberto:2023nym}, we compare results at the 14~TeV LHC and the nominal 100~TeV FCC.
Figure~\ref{fig:I} shows the rapidity-separation distribution, while Fig.~\ref{fig:pT} presents the Higgs transverse-momentum spectrum at fixed $\DY=3$.
The ancillary ratios isolate the impact of resummation and matching: the resummed-to-expanded comparison measures contributions generated beyond the fixed-order truncation, whereas ratios involving the matched result quantify the net effect of consistently combining the two descriptions.

The $\DY$ distributions exhibit an increasing separation between the NLL$^-$ resummed result and its fixed-order expansion as the rapidity interval grows.
This behavior is the expected imprint of BFKL dynamics, since larger rapidity separations enhance the energy logarithms resummed by the Green's function.
At the same time, the matched-to-resummed ratio tends toward unity in the large-$\DY$ region, indicating the progressive dominance of the high-energy component over the fixed-order baseline.
The transition is particularly visible at 14~TeV, while at 100~TeV the enlarged phase space distributes the effect over a broader kinematic domain.
A complementary picture emerges from the Higgs transverse-momentum distribution in Fig.~\ref{fig:pT}.
Around the spectral peak, matching produces sizeable corrections with respect to the fixed-order prediction, reaching approximately 30--50\%.
This pattern is consistent with the sensitivity previously identified in Ref.~\cite{Celiberto:2020tmb} to logarithmic contributions of the form $\ln(\hat{s}/p_H^2)$, which become enhanced when the Higgs boson and the recoiling jet carry comparable transverse momenta.
Toward larger transverse momentum, the matched uncertainty bands broaden and the pure high-energy description becomes progressively less constraining.
In this region, additional logarithmic structures, including collinear DGLAP-type enhancements and threshold effects~\cite{Bonciani:2003nt,deFlorian:2005fzc,Muselli:2017bad}, become increasingly relevant and are not fully controlled by energy resummation alone.
This behavior points to the need for complementary resummation mechanisms when approaching the high-transverse-momentum tail.
Importantly, the matched predictions reduce the tension between fixed-order and pure-HyF calculations previously observed at large transverse momentum, notably in Fig.~8 of Ref.~\cite{Celiberto:2020tmb}.
The POWHEG+JETHAD construction therefore provides a smoother interpolation between fixed-order and high-energy regimes while preserving their respective perturbative content.
This Higgs-plus-jet benchmark establishes the matching strategy that we subsequently extend to electroweak-boson plus jet production.

\section{From the Higgs benchmark to precision $Z$-plus-jet phenomenology}
\label{sec:roadmap}

The Higgs-plus-jet analysis provides a controlled testing ground for the matching of fixed-order perturbation theory with high-energy resummation.
Its main lesson is that energy-logarithmic effects can remain sizeable even for observables traditionally regarded as part of the precision-QCD domain.
Once a consistent POWHEG+JETHAD matching prescription is established, the same strategy can be transferred to electroweak final states for which direct experimental reconstruction offers additional information.
Drell-Yan production represents a particularly important target.
Its role as a standard candle for PDF determinations, electroweak precision measurements, and detector calibration implies that percent-level theoretical accuracy is required, especially at the High-Luminosity LHC.
Corrections at the ${\cal O}(10\%)$ level associated with high-energy dynamics therefore deserve a dedicated assessment rather than being treated as a secondary effect.
Among Drell-Yan channels, $Z$-plus-jet production is especially well suited to this purpose.
The leptonic decay of the $Z$ boson provides a fully reconstructible final state and gives access to angular distributions and polarization information, while the accompanying jet supplies the large rapidity separation needed to enhance high-energy logarithms.
This establishes a direct bridge between fixed-order, high-energy resummation theory, and realistic lepton-level measurements.
We therefore extend out matching strategy to a dedicated JETHAD-DYnamis--POWHEG framework for $Z$-plus-jet production.
The target is a resummation-improved description of rapidity and transverse-momentum spectra together with polarization-sensitive and angular observables, ultimately providing a precision framework for present LHC analyses and the High-Luminosity LHC.

\section*{Acknowledgments}
\label{sec:acknowledgments}

We are supported by the Atracci\'on de Talento Grant n. 2022-T1/TIC-24176 (Madrid, Spain).

\vspace{-0.05cm}
\begingroup
\setstretch{0.6}
\bibliographystyle{bibstyle}
\bibliography{bibliography}
\endgroup

\end{document}